\documentclass[a4paper,11pt]{article}
\usepackage{float}
\usepackage[pdftex]{graphicx}
\usepackage{subcaption}
\usepackage[T1]{fontenc}
\usepackage{lmodern}
\usepackage{slashed}
\usepackage[utf8]{inputenc}
\usepackage[english]{babel}
\usepackage{microtype}
\usepackage{cite}
\usepackage{amsmath,amssymb,amsfonts,amsthm}
\usepackage{mathtools,mathrsfs,calligra,aurical}
\usepackage[nottoc,notlot,notlof]{tocbibind}
\usepackage{upgreek}
\usepackage{mathtools}
\numberwithin{equation}{section}
\allowdisplaybreaks
\usepackage[all]{xy}
\usepackage{color}
\usepackage{graphicx}
\graphicspath{{images/}}
\usepackage{geometry}
\usepackage[toc,page]{appendix}
\usepackage{hyperref}
\usepackage[normalem]{ulem}

\usepackage{bm}
\usepackage{ragged2e}
\usepackage{appendix}
\usepackage{slashed}
\usepackage{bbold}
\usepackage{cancel}

\definecolor{blue-violet}{rgb}{0.54, 0.17, 0.89}
\definecolor{PineGreen}{cmyk}{0.92, 0, 0.59, 0.25}
\definecolor{YellowOrange}{cmyk}{0, 0.42, 1, 0}
\definecolor{orange}{rgb}{0.95, 0.5, 0.1}

\newcommand{\be}{\begin{equation}}
\newcommand{\bea}{\begin{eqnarray}}

\newcommand{\ee}{\end{equation}}
\newcommand{\eea}{\end{eqnarray}}

\def\a{\alpha}\def\b{\beta}

\DeclareMathAlphabet{\mathpzc}{OT1}{pzc}{m}{it}

\usepackage[dvipsnames]{xcolor}

\begin{document}

\begin{titlepage}
\begin{flushright}
\par\end{flushright}
\vskip 0.5cm
\begin{center}
\textbf{\LARGE \bf  Universal IR Holography, Scalar Fluctuations and Glueball spectra}\\
\vskip 5mm

\vskip 1cm

\large {\bf Andr\'{e}s Anabal\'{o}n}$^{~a ~b}$\footnote{anabalo@gmail.com} and \large {\bf Horatiu Nastase}$^{~b}$\footnote{horatiu.nastase@unesp.br}

\vskip .5cm

$^{(a)}${\textit{Departamento de Física, Universidad de Concepción,
Casilla 160-C, Concepción, Chile.
}}\\ \vskip .1cm 
$^{(b)}${\textit{Instituto de F\'isica Te\'orica, UNESP-Universidade Estadual Paulista \\
R. Dr. Bento T. Ferraz 271, Bl. II, Sao Paulo 01140-070, SP, Brazil.}}
\end{center}
\begin{abstract}
{We show that the d'Alembertian operator with a possible mass term in the AdS soliton and more general confining gravity dual 
backrounds admits infinitely many different 
spectra. These can be interpreted as different theories in the 
infrared and correspond to multitrace deformations of either the Dirichlet or the Neumann theory. We prove that all these 
fluctuations are normalizable and provide examples of their spectra. Therefore, the AdS soliton can be interpreted as 
giving a holographic RG flow between an universal UV theory at the AdS boundary and these infinitely many 
possibilities in the IR, obtained by deformations. 
The massive spectrum of the double trace deformation in $AdS_5$ allows the matching 
of the large-$N$ glueball masses of lattice $QCD_3$; the ratio of the ground states of the $2^{++}$ and $0^{++}$ 
channels are in full agreement with the lattice prediction. When considering $AdS_7$ and the 4-dimensional pure glue theory, 
a remarkably general picture emerges, 
where we can write formulas for the fluctuations that are in agreement with ones from holographic high-energy 
scattering and from AdS/CFT with IR and UV cut-off. We point out that this log branch in the IR in $D$-dimensions can 
be seen as the usual logarithmic branch of scalar fields saturating the Breitenlohner-Freedman bound in a conformally 
rescaled metric, with $AdS_{D-1}\times S^1$ asymptotics.}
\end{abstract}

\vfill{}
\vspace{1.5cm}
\end{titlepage}

\setcounter{footnote}{0}
\tableofcontents

\section{Introduction and Discussion}

Holography, in AdS/CFT and gauge/gravity duality, identifies the quantum states of supergravity in the bulk 
and those of the corresponding quantum field theory (QFT) on the boundary. These quantum states are 
required to be normalizable to belong  to the relevant Hilbert spaces, hence normalizable modes of the supergravity fields
are identified with the non-perturbative states in the QFT. However, a field can be logarithmically divergent and normalizable. The main objective of this letter is to analyze the rich physics that this logarithmic branch brings in.

Another, related, point we study is the generality of some simple formulas for the glueball spectra arising from gravity duals 
of $QCD_3$ (or rather, pure glue, $YM_3$) and $QCD_4$ ($YM_4$). 
The Witten model \cite{Witten:1998zw} for finite temperature AdS/CFT, 
arising from a certain $M\rightarrow \infty$ scaling of a black hole in $AdS_{D}$,  can be written in Euclidean space as 
\be
ds^2=\frac{r^2}{\ell^2}\left[d\tau^2 F(r)+d\vec{x}_{(D-2)}^2\right]+\ell^2\frac{dr^2}{r^2F(r)}\;,\;\; 
F(r)=1-\frac{r_0^{D-1}}{r^{D-1}}\;,\label{Wittenm}
\ee
and for $D=5$ ($AdS_5$), can be also obtained, once we add an extra $S^5$, as the near-horizon near-extremal D3-brane metric. 
If we compactify and reduce on the "Euclidean time" $\tau$, we get a model dual to pure glue in 3 dimensions, that we call
$QCD_3$. On the other hand, if we consider
near-horizon near-extremal D4-branes in Euclidean time, with 
\bea
ds^2&=&\left(\frac{U}{\ell}\right)^{3/2}\left(F(U)d\tau^2+d\vec{x}^2_{(4)}\right)+\left(\frac{\ell}{U}\right)^{3/2}\left(\frac{dU^2}{F(U)}+U^2
d\Omega_4^2\right)\cr
&=& 8\frac{\rho}{\ell}\left[\frac{\rho^2}{\ell^2}\left(F(\rho)d\tau^2+d\vec{x}^2_{(4)}\right)
+\ell^2\left(\frac{d\rho^2}{\rho^2F(\rho)}+d\Omega_4^2\right)\right]\;,\label{WSS}
\eea
where $\rho=(\ell U)^{1/2}/2$ and $F(U)=1-U^3_0/U^3$, so is conformal to asymptotically $AdS_6\times S^4$
(though, of course, the presence of the nontrivial, $U$-dependent dilaton $e^{\phi-\phi_0}=(U/\ell)^{3/4}$ means
that the solution itself does not have $AdS_6$ symmetries), and if we compactify
and reduce on $\tau$ we get a model dual to pure glue in 4 dimensions, that we call $QCD_4$ (or $YM_4$), 
that was expanded by the introduction of D8-branes by 
Sakai and Sugimoto \cite{Sakai:2004cn,Sakai:2005yt}. In both of these models, as well as in other holographic QCD 
constructions for ${\cal N}=1$ $SYM_4$ \cite{Klebanov:2000hb,Polchinski:2000uf,Maldacena:2000yy} and 
${\cal N}=1$ $SYM_3$ \cite{Maldacena:2001pb}, the metric in the IR is smooth, and locally equivalent to flat space, while 
in the UV it asymptotes to AdS space, perhaps log-corrected \cite{Klebanov:2000hb,Polchinski:2000uf}
or with an equivalent nontrivial dilaton 
\cite{Maldacena:2000yy,Maldacena:2001pb}, as needed for gravity duals of QCD-like theories (see, for instance, 
chapter 21 of \cite{Nastase:2015wjb} for more details). A simpler model for QCD is the "hard-wall" model \cite{Polchinski:2001tt}, 
where AdS space is cut-off in the IR, while the Randall-Sundrum model \cite{Randall:1999ee,Randall:1999vf} can also 
be understood as AdS/CFT with an extra UV cut-off, besides the IR cut-off. In all of these cases, the d'Alembertian operator
$\Box$ in the gravity dual, corresponding to glueball spectra, reduces to a one-dimensional Schr\"{o}dinger problem, and we 
will see that there are universal features, namely model independent, and how they reflect on the modes.

First, we will study a real scalar field in a general confining background. The discussion is quite generic, showing that always there 
is a solution with a logarithmic branch in the IR. Second, using the standard Klein-Gordon norm, we show that a scalar field 
with a logarithmic branch in the IR is indeed normalizable. Third, we show how to renormalize the action in the infrared and 
show how the logarithmic branch allows an holographic interpretation as a multi-trace deformations of the dual theory in the IR. 
Then we provide the spectra of several boundary conditions and we point out the relation of the log branch with that of a 
scalar field saturating the Breitenlohner-Freedman bound in $AdS_{D-1}\times S^1$.
Finally, we consider the generic behaviour of the UV and IR modes, from which we derive formulas for the mass spectrum 
that fit well the (lattice numerics) data, and show that they are compatible both with the 
KK modes of the Randall-Sundrum model \cite{Randall:1999ee,Randall:1999vf}, reinterpreted through AdS/CFT, and with the 
spectrum derived from
scattering of hadrons at high energy, via the Polchinski-Strassler scenario in the ``hard-wall'' model \cite{Polchinski:2001tt}, 
found and argued to be asymptotically {\em exact} in \cite{Kang:2004yk,Kang:2005bj,Nastase:2005bk,Nastase:2005rp}.

The paper is organized as follows. In section 2 we describe the IR logarithmic branch, its normalizability and its 
connection with the UV in $AdS_{D-1}\times S^1$, in section 3 we study applications to $QCD_3$ and $QCD_4$ models, 
with several numerical results and explicit details. Below, in the tables and plots of this work we give the masses in units of $1/z_0$.

\section{A logarithmic branch in the IR}

We consider the following gravity dual background, of which we see that the ``Witten model'', or ``AdS soliton'', 
is an example,\footnote{The condition for such a background to be dual to a confining theory can be found from 
\cite{Kinar:1998vq}.}
\begin{equation}
ds^2=\frac{\ell^2}{z^2}\left(F(z)d\theta^2+\frac{dz^2}{F(z)}+\gamma_{a b}dx^a dx^b\right)\, ,
\label{metric}
\end{equation}
where $\gamma_{a b}$ is a Lorentzian metric independent of $z$ and $\ell$ is the AdS radius, $z=\ell^2/r$ in (\ref{Wittenm}), and
$x^a=(t,\vec{x})$. 
We take $z\in[0,z_0]$. At the UV endpoint, $F(0)=1$, and at the IR endpoint, 
$F(z_0)=0$. Both are simple zeroes of $F$ and the function is 
otherwise positive. An important example in $D$-dimensions is $F=1-(z/z_0)^{D-1}$ and $\gamma_{a b}=\eta_{a b}$ is 
the Minkowski metric, which is the well-known ``Witten model'' or 
``AdS soliton'' \cite{Horowitz:1998ha}. It satisfies the Einstein equations 
with a negative cosmological constant $R_{\mu \nu}=-\frac{(D-1)}{\ell^2}g_{\mu \nu}$ and $\theta$ parameterizes a 
circle that has no conical singularities when $\theta\in[0,\theta_0]$. Other interesting and simple examples are the 
supersymmetric solitons of  \cite{Anabalon:2021tua}, which have $F=1-(z/z_0)^{4}$, for $D=4$, $F=1-(z/z_0)^{6}$, for $D=5$ and 
$\gamma_{a b}=\eta_{a b}$. Our discussion below on the existence and 
normalizability of the logarithmic mode in the IR depends only on the form of the metric in the neighborhood of $z_0$. 

Pick a real scalar field in this background with the standard, Lorentzian action
\begin{equation}
S_{\phi}=-\frac{1}{2}\int_{0}^{z_0}dz\int_{0}^{\theta_0} d\theta 
\int d^{D-2}x\sqrt{-g}\left(\left(\partial\phi \right)^2+m^2\phi^2\right)+S_{\partial}\;,
\label{action}
\end{equation}
where we include possible boundary terms, $S_{\partial}$, that will render the action principle well defined. 
The novelty is that $S_{\partial}$ arises in the IR; in the UV, we assume that the standard considerations apply, 
see the review \cite{Skenderis:2002wp} and references therein. 
Let us remark that in the IR, the procedure is not completely new: in the context of AdS/CFT in Minkowski space 
at finite temperature, 
Son and Starinets \cite{Son:2002sd} proposed to write boundary terms both in the UV and IR, and further Iqbal and Liu 
\cite{Iqbal:2008by} used a similar procedure to calculate transport coefficients at the horizon via the membrane paradigm;
this became a standard tool in AdS/CMT, see the review \cite{Hartnoll:2009sz} and the book \cite{Nastase:2017cxp} for more 
details. 
The wave equation is $\Box_g \phi -m^2\phi  =0$, 
when evaluated in (\ref{metric}) with the Frobenius ansatz $\phi=(z_0-z)^{\Delta}\Phi_M(t,\vec{x}) \cos(p\theta)$ and 
$\Box_{\gamma} \Phi_M -M^2\Phi_M =0$ yields $\Delta=\pm p\, c_0^{-1}$ under the assumption that 
the metric and the scalar field potential admits a Taylor series around $z_0$ of the form $F(z)=c_0(z_0-z)+O(z_0-z)^2$. It follows 
that when $p=0$ there is a logarithmic branch. This is the case of our interest. Hence, in what follows 
we restrict our study to $p=0$ and and the following expansion of the scalar field in the IR
\begin{equation}
\phi_i=\alpha_i(t,\vec{x})\sum^{\infty}_{n=0}a_n\left(1-\frac{z}{z_0}\right)^n
+\beta_i(t,\vec{x})\ln\left(1-\frac{z}{z_0}\right)\sum^{\infty}_{n=0}b_n\left(1-\frac{z}{z_0}\right)^n\;,
\label{exp}
\end{equation}
where we fix the redundancy in the parameterization setting $a_0=b_0=1$. Moreover, this logarithmic 
branch is generic for all solitons, in all theories, in all dimensions provided that the background admits 
this expansion; for several examples see  \cite{Anabalon:2019tcy, Anabalon:2021tua, Anabalon:2022ksf, Anabalon:2022aig, Anabalon:2023oge,Quijada:2023fkc, Nunez:2023nnl}. 

The Klein-Gordon current in $D$-dimensions is $J_{\mu}=-i \left(\phi_1^*\partial_{\mu} \phi_2-\phi_2\partial_{\mu} 
\phi_1^*\right)$, where $\phi_i$ are two different solutions to the Klein-Gordon equation. For a detailed discussion of normalizability in AdS/CFT see for instance  \cite{Andrade:2011dg}. Let us prove that the logarithmic mode is normalizable with this 
norm.\footnote{This is the norm that is finite and conserved in time for solutions with arbitrary time dependence of the wave equation; otherwise other 
norms could be imagined. We thank Carlos N\'{u}\~nez for a discussion on this point.}
To do this we first show that is possible to 
define a conserved and finite norm from $J^\mu$ independently of $z$ including the logarithmic branch. 
To pass from the current to the norm it is useful to note that, because the fluctuation of interest is independent of $\theta$, $J_{\theta}=0$, and 
\begin{equation}
\int_{0}^{z_0}dz\sqrt{-g} \nabla_{z}(J^z) =J_z \, g^{z z}\sqrt{-g}\Big|_0^{z_0}=i c_0\left(\beta_2 \alpha^*_1
-\beta^*_1 \alpha_2\right) \frac{\ell^{D-2}}{z_0^{D-2}}-UV\;.
\label{IR}
\end{equation}

We suppose that the UV term vanishes as the fluctuations are normalizable there. The IR term vanishes with 
Dirichlet boundary conditions $\alpha=0$, Neumann boundary conditions $\beta=0$, or double trace boundary 
conditions $\beta=\lambda \alpha$. We will justify the denomination ``double-trace'' below. 
With these IR conditions the constant time integral of $\nabla_{\mu}J^{\mu}=0$ yields a conserved norm
\begin{equation}
0=\int dzd^{D-3}x \sqrt{-g}\nabla_{\mu}J^\mu =\frac{d}{dt}\int dzd^{D-3}x g^{t t}J_t\sqrt{-g}\;.
\label{norm}
\end{equation}

We note that for black holes (in Lorentzian signature!) $g^{t t}$ ($\propto F(z)^{-1}$) is singular around $F(z)= 0$, 
whereas for solitons 
($=z^2/l^2$) it is regular in the IR, which allows for a finite result. 
Therefore, the $z$-integrals are manifestly finite in the IR, as follows from the expansion (\ref{exp}). 
Also note that, while scalars of interest might be real, the standard prescription for the KK soliton (here AdS soliton), as 
in \cite{Witten:1981gj,Horowitz:1998ha,Anabalon:2021tua}, allows for discarding a part of the scalar field (with negative energy), 
and thus keeping a complex part.

In order to define a standard QFT, the states with positive energy should have positive norm. 
It should be clear from the previous discussion that a solution with a well defined mass $\phi=\Phi_M(t,\vec{x})X_M(z)$ 
has a positive norm whenever they have a positive energy, $\partial_t\Phi_M=-i\omega\Phi_M$. 
Indeed, in this case the conserved norm (\ref{norm}) reduces to the standard Klein-Gordon norm in flat space for $\Phi_M$ 
times a manifestly positive integral,
\begin{equation}
N_M\equiv-\int_0 ^{z_0} dz\sqrt{-g} g^{t t} X_M(z)^2>0\;.
\end{equation}

The variation of the action (\ref{action}) with these new boundary conditions gets a contribution in the IR.  
Hence, counterterms need to be included there, a divergent and a finite one,
\begin{equation}
S_{\partial}=-4\pi\frac{\ell^{D-2}}{z_0^{D-2}} \int d^{D-2}x\sqrt{-\gamma}\left(\frac{\beta^2}{2}\ln{\epsilon}+S_F\right)\;,
\end{equation}
where $\epsilon$ is an IR regulator defined as $z=z_0-\epsilon z_0$, such that the IR is located at $\epsilon=0$, and 
$S_F$ is a finite counterterm that one can add and that changes the IR theory, where we can have, for instance,
\bea
S_F=0\Longrightarrow \delta S_\phi &=&4\pi\frac{\ell^{D-2}}{z_0^{D-2}}\;  \int d^{D-2}x\sqrt{-\gamma} \bm{\beta \delta \alpha}\;, \cr
S_F=\alpha\beta\Longrightarrow \delta S_\phi&=&{\bm -}4\pi\frac{\ell^{D-2}}{z_0^{D-2}}\;\int d^{D-2}x\sqrt{-\gamma}  
\bm{ \alpha \delta \beta} \;, \cr
S_F=\frac{\beta^2}{2\lambda}\Longrightarrow  \delta S_\phi
&=&4\pi\frac{\ell^{D-2}}{z_0^{D-2}}\; \int d^{D-2}x\sqrt{-\gamma}\bm{ \beta(\delta \a-\delta \b /\lambda)}\cr
&\equiv&
4\pi\frac{\ell^{D-2}}{z_0^{D-2}}\; \int d^{D-2}x\sqrt{-\gamma}\beta \delta J_\beta   \;, \cr
S_F=\alpha\beta-\frac{\lambda\alpha^2}{2}\Longrightarrow  \delta S_\phi
&=&{\bm -}4\pi\frac{\ell^{D-2}}{z_0^{D-2}}\;\int d^{D-2}x\sqrt{-\gamma}  \bm{ \a(\delta\b -\lambda\delta \a)}\cr
&\equiv&
-4\pi\frac{\ell^{D-2}}{z_0^{D-2}}\;\int d^{D-2}x\sqrt{-\gamma} 
\alpha \delta J_\alpha\;,\label{doublebeta}
\eea
where $J_\beta=\alpha-\frac{\beta}{\lambda}$ and $J_\alpha=\beta-\lambda \alpha$ are generalized 
sources and $\lambda$ is a constant. As the renormalized action is finite 
one should consider $\lambda$ as a {\em renormalized} coupling constant. One can see that the possibilities in 
the last two cases, with $S_F=S_F(\lambda)$, are double 
trace deformations of the dual theory in the IR, in the sense of \cite{Witten:2001ua}. 

It is interesting to note that for the 
third case in (\ref{doublebeta}), with source $J_\b=\a-\lambda^{-1}\b$ and deformation $S_F=\lambda^{-1}\b^2/2$, 
the coupling constants are related by 
$\frac{1}{\lambda} + \ln{\epsilon} =\frac{1}{\lambda_0}$,
where $\lambda_0$ is the bare coupling constant. Let us set $\epsilon=\frac{\mu}{\Lambda}$, with 
$\Lambda\gg\mu$\;\;\footnote{
In this language it is natural to think of the IR scale of the theory as $\Lambda=z_0$ and the renormalization scale 
as $\mu=z_0-z$ with fixed $z$.}, then one can recover the typical relation between the renormalized coupling and 
the bare coupling in a theory with a beta function running only at one-loop,
\begin{equation}
\lambda=\frac{\lambda_0}{1+\lambda_0\ln{\frac{\Lambda}{\mu}}}.
\label{loop}
\end{equation}

This discussion is the same as the one given in \cite{Witten:2001ua}, giving the same (\ref{loop}), but there in the UV. 
However, in \cite{Witten:2001ua},
${\lambda}$ is in the numerator of $S_F$ and ${\Lambda}$ is a UV cutoff. Here ${\lambda}$ is in the denominator of 
the deformation $S_F$, and we have an {\em IR cutoff}, which is ${\Lambda_{\rm QCD}}$ in this holographic context, 
so it is natural to replace $1/\epsilon=\Lambda/\mu$ and $\lambda=\lambda(\Lambda)$, with 
$1/\epsilon=\mu/\Lambda_{\rm QCD}$ and $\lambda=\lambda(\mu)$, as noted. The fact that the coupling 
constant appears in the denominator of $S_F$ makes it natural to identify ${\lambda}=\pm g^2_{\rm QFT}$, where $g^2_{\rm QFT}$ 
is the standard coupling constant of a non-abelian gauge theory, with the beta function
\begin{equation}
\mu\frac{\partial g_{\rm QFT}}{\partial\mu}=-\frac{\pm 1}{2}g^3_{\rm QFT}\;.
\end{equation}

It is not completely clear what is the exact interpretation of the formula
(\ref{loop}) in field theory, though one possible interpretation would be that there 
is an effective quantum IR description, according to the rules above, 
that is one-loop exact (since presumably, higher loop corrections would modify the formula). 

In conclusion, we replace an UV renormalization with an IR renormalization, that at one loop still looks formally the same.
In these double trace deformations, (both $J_\b=0$ and $J_\a=0$ imply that) we have $\a=\lambda^{-1}\b$, as an {\em on-shell}
relation (at zero source). We notice that the behaviour is that of asymptotic freedom in the IR for $\lambda>0$ and the theory 
is strongly coupled in the IR for $\lambda<0$. We have chosen our conventions such that the sign of $\lambda$ indicates 
the sign of the contribution to the energy. Hence, the behaviour of the fluctuations with positive or negative $\lambda$ 
should be very different. This will be shown to be the case in what follows.

\subsection{Conformal relation to $AdS_{D-1}\times S^1$}

The IR structure we propose here is analogous to what happens at
the usual AdS boundary.  Therefore, one might wonder whether the IR is related to a geometry with an AdS factor. 
This is indeed the case, as we will show below. Furthermore, we will see how to map the dynamics of the fluctuations 
in one metric to the other in the neighborhood of $z=z_0$. 

First, note that the conformally rescaled metric (\ref{metric}),
\begin{equation}
ds_{AdS_{D-1}\times S^1}^2=\frac{ds_g^2}{F(z)}=\frac{\ell^2}{z^2}\left(d\theta^2+\frac{dz^2}{F(z)^2}
+\frac{1}{F(z)}\gamma_{a b}dx^a dx^b\right)\, ,
\label{metric2}
\end{equation}
is asymptotically $AdS_{D-1}\times S^1$. This can be easily seen by going to the neighborhood of $z=z_0$,
 setting $F(z)=c_0 (z_0-z)$, and changing the coordinates as
\begin{equation}
z=-\frac{1}{4}c_0 Z^2+z_0 \, ,
\end{equation}
which shows that (\ref{metric2}) is asymptotically $AdS_{D-1}\times S^1$ around $Z=0$. The radius of this 
new $AdS_{D-1}$ is $L=\frac{2 \ell}{z_0 c_0}$, which is also the radius of the $S^1$. Hence, we find that when $F(z)=c_0 (z_0-z)$ we have
\begin{equation}
ds_{AdS_{D-1}\times S^1}^2\simeq L^2 d\tilde{\theta}^2+\frac{L^2}{Z^2}\left(dZ^2+\gamma_{a b}dx^a dx^b\right)\, .
\label{metric3}
\end{equation}
with $\tilde{\theta}\in [0,2\pi)$.
For fluctuations independent of the angle $\theta$ or equivalently of $\tilde{\theta}$, we find that 
\begin{equation}
F(z)^{\frac{2-D}{4}}\left(\Box_{AdS_{D-1}\times S^1} \psi -\mu^2\psi\right)-F(z) \left(\Box_g \phi -m^2\phi\right)=O(z_0-z)\;,
\label{map}
\end{equation}
provided that $\psi=\phi \, F(z)^{\frac{D-2}{4}}$, and that the mass of the fluctuation $\psi$ in the conformally 
rescaled metric saturates the Breitenlohner-Freedman bound in that space, $\mu^2=-\frac{(D-2)^2}{4 L^2}$ \footnote{Note 
that in (\ref{map}) the $m^2$ term becomes subleading in this IR region of the original metric, and instead of it, we 
now have $\mu^2$, leading in (the UV of) $AdS_{D-1}\times S^1$.}. This observation 
seems interesting, since it might help to export holographic techniques and concepts to other space-times. In particular, in 
this case we see that the log-branch we have discussed so far is, asymptotically, the usual logarithmic branch of a scalar 
field saturating the Breithenloner-Freedman bound in $AdS_{D-1}\times S^1$. 

This discussion it is instructive to see that from the point of view of the $AdS_{D-1}\times S^1$, the $\alpha$ and  
$\beta$ of (\ref{exp}) have the same conformal dimension, $\Delta$, corresponding to a scalar field saturating the 
Breithenloner-Freedman bound, $\psi$, namely $\Delta=\frac{D-2}{2}$. From the point of view of the original metric, 
and scalar field $\phi=\psi\, Z^{-\frac{D-2}{2}}$, it follows that $\alpha$ and $\beta$ are invariant under rescaling in $Z$. 
Therefore the deformations discussed above, (\ref{doublebeta}), can be thought of as relevant, with $Z$ as the energy scale. 
Then a particular IR theory corresponds to a particular value for these $\a$ and $\b$.

\section{Applications}

\subsection{Massless modes and validity of the probe limit}

Our previous discussion has been rather general, now we pass to discuss concrete theories. The first important issue 
would be whether these modes can be treated as probes on a geometry as they can potentially deform it due to its 
logarithmic IR behaviour. This turn out to be intrinsically related to the existence of a possible massless mode for 
certain deformations. It is relatively simple to construct such mode in the case of the AdS Soliton for a scalar field 
of the form $\Box\Phi=0$, the massless mode is $\Phi=\Phi_0(t,\vec{x}) \ln \left(1-\left(\frac{z}{z_0}\right)^{D-1}\right)$, 
with $\eta^{a b}\partial_a\partial_b\Phi_0=0$. This mode is indeed normalizable, since the norm $N_M=N_0$ satisfies
\begin{equation}
N_0\leq \zeta(3)\;,
\label{masless}
\end{equation}
with $\zeta$ the Riemann zeta function and equality saturated for $D=3$. This massless mode has IR boundary 
conditions $\a=\b\ln(D-1)$. However, whenever this massless mode exists there is also a zero mode that 
changes the geometry. The most general static solutions of the Einstein-massless scalar system with a cosmological 
constant are the naked singularities found in \cite{Saenz:2012ga}. When the scalar field is normalizable in 
the UV, all these solutions have the IR boundary conditions 
$\a=\b\ln(D-1)$. \footnote{To show this one has to put the metric given in eq. (13) of \cite{Saenz:2012ga} in the gauge 
where $g_{xx}=g_{zz}$; $g_{xx}$ is the metric used in \cite{Saenz:2012ga} and $g_{zz}$ is our metric (\ref{metric}). 
This yields the change of coordinates $x=b\frac{2\, z_0^{D-1}-z^{D-1}}{z^{D-1}}$.} These pathological cases are 
therefore excluded if the boundary condition of the gravity theory 
is such that $\lambda^{-1}\neq\ln(D-1)$. Hence, when the probe limit is valid these new IR boundary conditions exclude 
massless modes on the brane.\footnote{The logarithmic branch was previously excluded in its entirety, without taking into 
account these subtleties, see e.g. \cite{deMelloKoch:1998vqw}.} This can be understood as a no-hair theorem for massless 
scalar fields in AdS. Namely, in the Einstein-massless Klein Gordon-AdS system the only solution that is normalizable 
in the UV satisfies  $\a=\b\ln(D-1)$ in the IR. Therefore if the case $\a=\b\ln(D-1)$ is excluded, the only static solution 
is that of pure General Relativity with cosmological constant. Hence, the probe limit is valid provided on-shell the scalar 
field satisfies $\lambda^{-1}\neq\ln(D-1)$ and is normalizable in the UV. 

\subsection{$QCD_{3}$}

Further insight can be gained by going to a concrete dimension. Let us focus on $D=5$ and type IIB supergravity. 
The deformation has $\a=\lambda^{-1}\b$.
In this case the massless scalar field can be identified with the dilaton of string theory. The glueballs  
dual to the operator ${\rm Tr}\;  F_{\mu\nu}^2$,
are in the $0^{++}$ channel. A remarkable qualitative feature of the spectra of the new boundary conditions is 
that the $2^{++}$ spectrum (given by the $\beta=0$ column of table 1, or $\lambda^{-1}=\pm\infty$) 
is no longer degenerate with that of the 
$0^{++}$ one generated by the dilaton. This is rather desirable, since to break this degeneracy was an open problem, 
in the attempts to describe the glueball spectrum of QCD with holographic techniques. Furthermore, the lightest 
glueball in the $0^{++}$ channel can be seen as arising from the dilaton (previously it was from the metric, see 
\cite{Brower:2000rp}). The double trace deformation allows us to pick $\lambda$, in order to match the mass ratio of the 
ground states of some arbitrary desired channel. For instance, in the large $N$ limit, lattice $QCD_3$ predicts a 
ratio for the ground states $M_{2^{++}}/M_{0^{++}} \approx 1.68$ \cite{Athenodorou:2016ebg}. This is exactly 
reproduced in the gravity side with a ground state of $M_{0^{++}}^2\approx4.1/z_0^2$; namely $\lambda\approx e$. 
In other words, choosing the background (with a given $\ell$ and $z_0$), fixes $M_{2^{++}}$, and fixing $\lambda\approx e$ 
then achieves the best fit. 
This is, so far, the best quantitative matching of top-down holography to lattice QCD in the literature\footnote{See, for instance, 
\cite{Sonnenschein:2018fph,Dymarsky:2022ecr}. Note also \cite{Sil:2017lbb} for a top-down construction for $QCD_4$.}, 
and is achieved 
by a nontrivial choice of $\lambda$. This is a rather unexpected result as lattice QCD calculations are at weak t'Hooft 
coupling and the gravity result is at strong t'Hooft coupling\cite{Caceres:2005yx} \footnote{We thank Carlos N\'{u}\~nez 
for pointing out this 
important detail to us.}. 

\begin{table}
\centering
\begin{tabular}{ |p{3cm}|p{3cm}|p{3cm}|  }
 \hline
 \multicolumn{3}{|c|}{Mass$^2$ spectrum for different IR boundary conditions for $QCD_3$} \\
 \hline
 \hline
 $\beta=0$&$\alpha=0$&$\alpha=2\beta$\\
 \hline
11.5878&5.05077&3.76515\\
34.537&27.4248&26.6285\\
68.9753&60.424&59.5767\\
114.987&104.961&103.975\\
 \hline
\end{tabular}
\newline
\caption{\label{masses} Masses$^2$ of the $0^{++}$ glueballs from the dilaton in type IIB supergravity. The $\beta=0$ 
column coincides with the numerics available in the literature \cite{Csaki:1998qr, Brower:2000rp}, the other spectra are 
given as possible examples of the formalism developed here and are new.}
\end{table}

\begin{figure}
\centering
\includegraphics[scale=1]{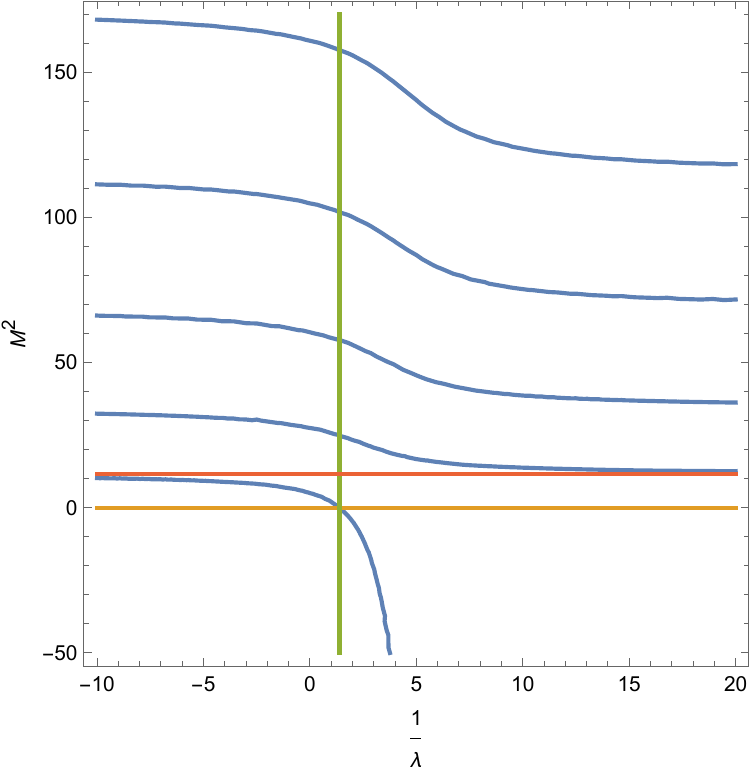}
\caption{M$^2$ vs $\lambda^{-1}$ for the multitrace deformation $\alpha=\lambda^{-1} \beta$ for $QCD_3$. 
Glueball mass$^2$ as a function 
of the inverse of the coupling constant $\lambda$. The green line is $\lambda^{-1}=\ln{4}$. For $\lambda^{-1}>\ln{4}$ there is a tachyon. 
At $\lim_{\lambda\rightarrow 0^{-}}M^2$ the Neumann theory is recovered. The Dirichlet theory is located at $\lambda^{-1}=0$. We 
observe that $M_{n}^2(\lambda^{-1}=-\infty)=M_{n+1}^2(\lambda^{-1}=\infty)$ for each eigenvalue labeled by $n$. For the ease of visualization of this property we include the red line which shows the matching of the asymptotic values of the lowest eigenvalue.}
\label{fig1}
\end{figure}

In Fig.\ref{fig1} we show how the value $\lambda^{-1}=\ln{4}$ is the onset of an instability characterized by 
the existence of tachyonic glueballs in the region $0<\lambda<\frac{1}{\ln{4}}$. The Neumann theory ($\b=0$) is located at $\lim_{\lambda\rightarrow 0^{-}}M^2$. Remarkably enough, the spectrum shows the property that $M_{n}^2(\lambda^{-1}=-\infty)=M_{n+1}^2(\lambda^{-1}=\infty)$ for each eigenvalue labeled by $n$.

\subsubsection{Conformal Quantum Mechanics}

Conformal quantum mechanics is characterized by a potential of the form $1/u^2$, see \cite {deAlfaro:1976vlx}. The D'Alembertian operator in the AdS soliton background can be seen to have this conformal behaviour in the IR and in the UV. 
If we replace the ansatz $\Phi=e^{-i(\omega t-\vec{k}\cdot\vec{x})}\frac{z^{3/2}}{F(z)^{1/4}}\Psi(u)$, with 
$\frac{dz}{du}=\sqrt{F(z)}$, in the wave equation $\Box \Phi=0$ we get (through the usual procedure) the one-dimensional 
Schr\"{o}dinger problem
\bea
-\frac{d^2\Psi(u)}{du^2}+V(u)\Psi(u)&=&(\omega^2-k^2)\Psi(u)\equiv M^2\Psi(u)\;,\cr
V(z)&=&-z^{5/2}F(z)^{1/4}\frac{d}{dz}\left(\frac{F(z)}{z^3}\frac{d}{dz}\left(\frac{z^{3/2}}{F(z)^{1/4}}\right)\right)\;.
\eea

For the AdS soliton in five dimensions one can actually find the change of coordinates from $z$ to $u$ explicitly, from $dz/du$. 
We have 
$z=z_0 \; sn(\frac{u}{z_0},i)$, which is the Jacobi elliptic sine of modulus $i$. This means that the interval over 
which the spacetime is defined is mapped as $z\in(0,z_0)\rightarrow u\in(0,z_0 K)$, where $K \approx 1.31102$ is the 
complete elliptic integral at modulus $i$. We observe the same power-law behaviour in the UV and the IR of $V(u)$, namely
\begin{align}
V(u\simeq 0)&\approx \frac{15}{4 u^2}+O(u^6)\;,\\
V(u\simeq K z_0)&\approx \frac{-1}{4(u-K z_0)^2}+\frac{4}{z^2_0}+O((u-K z_0)^2)\;.\label{IRpot}
\end{align}

The behaviour around these two points is that of conformal quantum mechanics, but with different coupling constants, 
namely $15/14$ in the UV and $-1/4$ in the IR. The value of $-1/4$ is the {\em analogue of the saturation of the 
Breitenlohner-Freedman bound} for the conformal quantum mechanics describing the gravity dual by reduction to 
$0+1$ dimensions, and generates the log branch in the IR.  The IR solution, using only the divergent part of the potential, 
is 
\be
\Psi=\sqrt{Kz_0-u}[C_1 J_0((u-Kz_0)M)+C_2 Y_0((u-Kz_0)M)]\;,\label{JYapprox}
\ee
with $J_0$ and $Y_0$ the standard Bessel functions of degree zero. 
The Neumann boundary condition sets $C_2=0$ (eliminating the log-divergent mode $Y_0$), 
and the UV boundary condition, {\em imposed assuming that the IR solution is valid throughout}, an 
approximation that could be questioned, but will be justified {\em a posteriori} shortly, implies then  the quantization 
condition, giving the masses of the $0^{++}$ glueballs, 
\be
J_0(K z_0 M_n)=0\;.\label{J0quant}
\ee

\begin{figure}[H]
\centering
\includegraphics[scale=0.5]{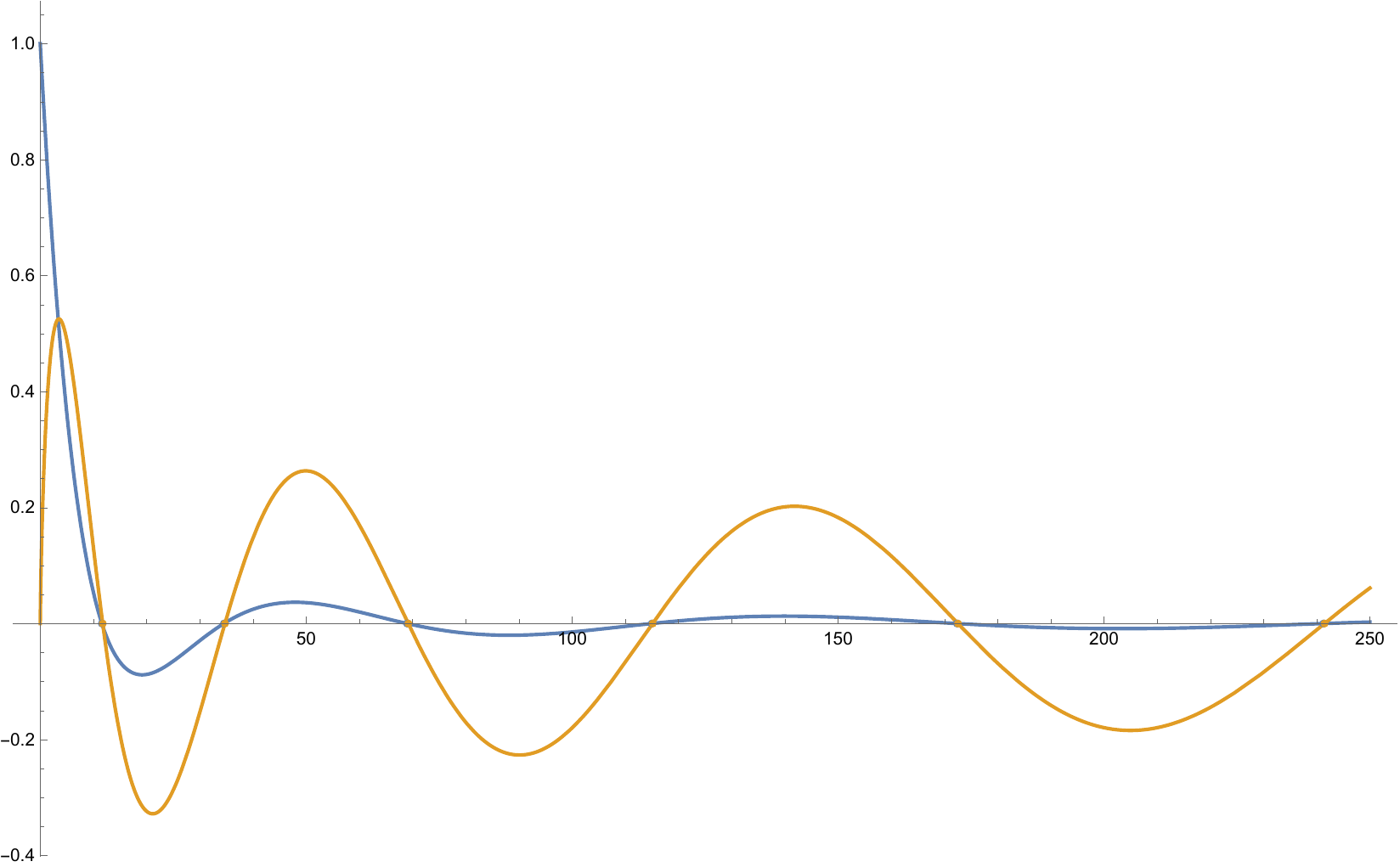}
\caption{The numerical solution for the D'Alembertian operator on the AdS Soliton in blue vs $J_{3/2}(K\, M)$ in orange. One can see that the the zeroes of both functions are in very good agreement. The $x$ axis is $M^2$.}
\label{figJ}
\end{figure}

The matching of the exact numerical solution with Dirichlet boundary condition to the IR solution (\ref{JYapprox}), exact
for small argument of $J_0$,
is not too good, but for its zeroes it is quite good; the match to the Bessel function $J_{3/2}$ is 
actually almost perfect as we see in Fig.\ref{figJ}. 

Note that in the UV, where the space is just AdS, we have the same solution for $\Psi$, just replacing $J_0,Y_0$ with 
$J_2,Y_2$ (and this corresponds to large argument for $J_2,Y_2$). But since 
$J_2(t)+J_0(t)=\frac{2}{t}J_1(t)$,  $Y_2(t)+Y_0(t)=\frac{2}{t}Y_1(t)$,
and at $|t|\rightarrow\infty$, 
$J_{3/2}(t)\simeq \frac{J_1(t)+Y_1(t)}{\sqrt{2}}=\frac{t}{2\sqrt{2}}(J_0(t)+Y_0(t)+J_2(t)+Y_0(t))$,
this suggests that the average of the IR and UV solutions, with the IR given by 
the sum of the constant mode $J_0$ and the log-divergent, 
yet normalizable one, $Y_0$, provides a better fit to the exact numerical solution with Neumann boundary conditions, 
whose zeroes give the quantization condition, replacing (\ref{J0quant}), thus the zeroes of $J_{3/2}$, or $J_1+Y_1$
replace the zeroes of $J_0$. 

We note that the application of the $AdS_5$ soliton to the holographic superconductor in \cite{Nishioka:2009zj} 
gave similar results to our glueball spectrum, for the nontrivial deformation of the conductivity 
$\sigma(\omega)$ by the analogue of the 
$\lambda$ deformation we considered (see, for instance, their Fig.3 for the behaviour in the UV), since 
$1/\lambda$ takes the role of the conductivity $\sigma$, and $M^2$ of $\omega$. The holographic superconductor is a scalar 
deformation of the horizon (the IR theory), just that one usually calculates transport coefficients 
in the UV; but via Iqbal and Liu's membrane paradigm \cite{Iqbal:2008by}, that is equivalent with an IR calculation at the horizon
under certain conditions.

\subsection{$AdS_7$ and 4-dimensional pure glue theory}

\begin{figure}[H]
\centering
\includegraphics[scale=1]{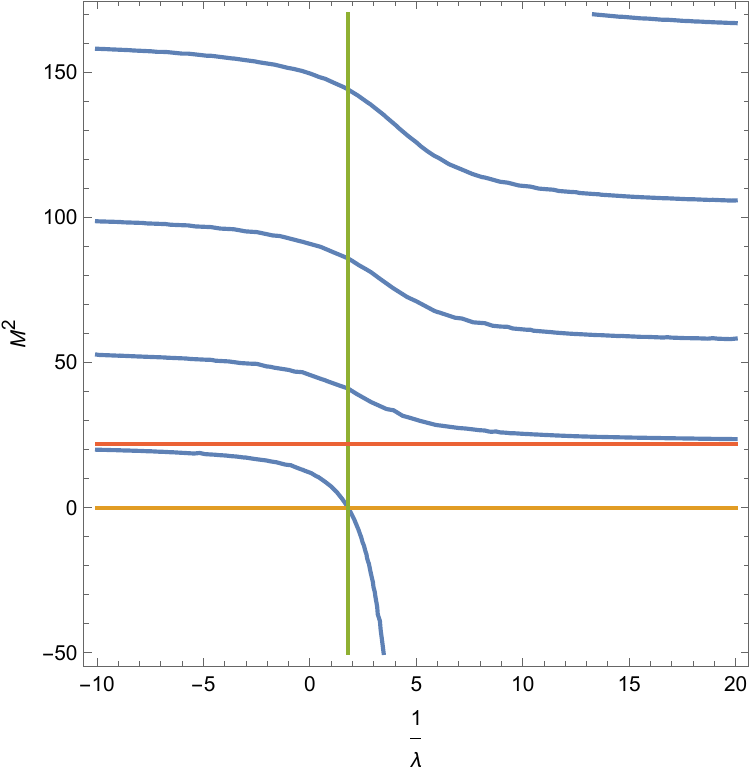}
\caption{M$^2$ vs $\lambda^{-1}$ for the multitrace deformation $\alpha=\lambda^{-1} \beta$ for the AdS soliton in $D=7$. The green line is $\lambda=\ln{6}$. For $\lambda^{-1}>\ln{6}$ there are tachyons. 
At $\lambda^{-1}=0$ the Neumann theory. We 
observe again that $M_{n}^2(\lambda^{-1}=-\infty)=M_{n+1}^2(\lambda^{-1}=\infty)$ for each eigenvalue labeled by $n$. The red line ease the visualization of this property.}
\label{fig2}
\end{figure}

In this case the fluctuations of the different supergravity fields have to be combined with other metric fluctuations to yield 
the dual of the $0^{++}$ glueball spectra of the pure glue theory. 
However, by comparing with \cite{Brower:2000rp} we find that that the 
operator associated to $T_4(r)$ in \cite{Brower:2000rp} is the same as that of the D'Alembertian of the massless scalar. 
To make the logarithmic branch relevant for the glueballs in this case, one should verify that it is normalizable with the 
norm associated to metric fluctuations. Hence we study this scalar fluctuation, normalizable with respect to the Klein-Gordon 
norm, but we postpone its connection to glueballs for a future work.   
Here the relevant background is the soliton in $AdS_7\times S^4$, corresponding to near-horizon near-extremal 
M5-branes, after we compactify two directions to get effectively $D=4$ on the boundary, and thus obtaining 
the 4-dimensional pure glue ($QCD_4$) theory. Equivalently, consider the 
Witten-Sakai-Sugimoto set-up (\ref{WSS}), of near-horizon near-extremal D4-branes, with background metric
conformal to $AdS_6\times S^4$.
The discussion is qualitatively the same. 
The spectrum can be found in table 2, and Fig.\ref{fig2} shows the masses$^2$ and deformations with $\lambda$.

For the solution of the $\Box\Phi=0$, in a general dimension $D$ of the AdS soliton, we have the ansatz
$\Phi=e^{-i(\omega t-\vec{k}\cdot\vec{x})}\frac{z^{(D-2)/2}}{F(z)^{1/4}}\Psi(u)$, again with 
$\frac{dz}{du}=\sqrt{F(z)}$, and for the Schr\"{o}dinger problem we get 
\be
V(z)=-z^{\frac{D-2}{2}}F(z)^{1/4}\frac{d}{dz}\left(\frac{F(z)}{z^{D-2}}\frac{d}{dz}\left(\frac{z^{\frac{D-2}{2}}}{F(z)^{1/4}}\right)\right)\;,
\ee
which gives the same IR potential (\ref{IRpot}), so the same IR solution (\ref{JYapprox}), but the UV potential is 
now $V(u\simeq 0)\approx \frac{D(D-2)}{4u^2}+O (u^6)$, so the UV solution is 
\be
\Psi=\sqrt{u}\left[C_1 J_{\frac{D-1}{2}}(uM)+C_2 Y_{\frac{D-1}{2}}(uM)\right]\;.
\ee

In the case relevant here, of $D=7$, we get the functions $J_3$ and $Y_3$ in the UV, instead of $J_2$ and $Y_2$ as in the
$QCD_3$ case. But since $J_3(t)=(4/t)J_2(t)-J_1(t)$, $Y_3(t)=(4/t)Y_2(t)-Y_1(t)$, at least at large $t$ (large $z_0M$), 
if we take a combination of the IR and UV solutions, again a linear combination of the $J_1$ and $Y_1$ gives the correct 
quantization condition.

\begin{table}
\centering
\begin{tabular}{ |p{3cm}|p{3cm}|p{3cm}|  }
 \hline
 \multicolumn{3}{|c|}{Mass$^2$ spectrum for different IR boundary conditions for $AdS_7$} \\
 \hline
 \hline
 $\beta=0$&$\alpha=0$&$\alpha=-\beta$\\
 \hline
22.1007&11.9994&14.5728\\
55.5851&45.5793&47.3567\\
102.452&90.9102&92.7819\\
162.708&149.552&151.554\\
 \hline
\end{tabular}
\newline
\caption{\label{masses 2} Masses$^2$ for the scalar fluctuations on the AdS soliton in $D=7$, corresponding to 
4-dimensional pure-glue theory. The $\beta=0$ 
column coincides with the numerics available in the literature for the T4 in \cite{Brower:2000rp}, the other spectra are new.}
\end{table}

\subsection{Universal quantization relations}

We have seen that the quantization conditions for the mass of the $QCD_3$ glueballs were, in a first approximation, 
written in terms of the zeroes of $J_0$, $j_{0,n}$, as in (\ref{J0quant}). In a better approximation, this was given by the 
zeroes of $J_1$, or rather of $J_1+Y_1$. These $M_n$'s for glueball states were obtained from the $\Box$ operator 
eigenstates in the AdS soliton, with an essential contribution from the IR of the soliton solution. 

But there are other two cases that result in masses for glueball solutions. One is the case of high-energy hadron scattering in the Polchinski-Strassler scenario, considered in the limiting case that 
saturates the Froissart unitarity bound, analyzed in \cite{Kang:2004yk,Kang:2005bj,Nastase:2005bk,Nastase:2005rp}. 
There it was argued that the gravity dual matches exactly the Heisenberg model for high-energy scattering in field theory, 
when at sufficiently high energies the scattering looks like a collision of gravitational shockwaves in the gravity dual, creating 
a black hole. In the limiting (Froissart saturated) case, the gravity scattering happens effectively on the IR cut-off (IR brane), 
and produces a black hole entirely on this IR cut-off and completely classical (with negligible fluctuations). The gravitational 
shockwave profile in AdS with this IR cut-off was calculated in \cite{Kang:2004yk} and found to be, at the IR cut-off position $y=0$,
\be
f(r)=\Phi(r,y=0)=R_s\sqrt{\frac{2\pi \ell}{r}}\sum_{n\geq 1}\frac{j_{1,n}^{-1/2}J_2(j_{1,n})}{a_{1,n}}e^{-M_n r}\;,\;\;\;
M_n=\frac{j_{1,n}}{\ell}\;,
\ee
where $M_n$ are the glueball states, $j_{1,n}$ are the zeroes of $J_1$,
$\ell$ is the AdS radius and $R_s\sim G_4\sqrt{s}$ the Schwarzschild radius of the shockwave collision. 
Thus, as usual, the states in the IR dominate the high-energy scattering giving the Froissart bound: keeping only $M_1$, then 
the black hole horizon radius $r_H$ is reached when $f(r)\sim 1$, giving a Froissart cross section $\sigma(s)\sim \pi r_H^2\sim 
\frac{\pi}{M_1^2}\ln ^2(\sqrt{s} G_4 M_1)$ \cite{Kang:2005bj}.

So the glueball states in this case are given by the quantization condition $J_1(\ell M_n)=0$. 

We can also obtain the glueball states by considering the Randall-Sundrum model
\cite{Randall:1999ee,Randall:1999vf}, understood as $AdS_5/CFT_4$ 
with both an IR and an UV cut-off. Then the solution for the graviton wavefunction at small $M$ is of the $J_2+Y_2$ type,
\be
\psi=N_M (|z|+1/k)^{1/2}\left[Y_2(M(|z|+1/k))+\frac{4k^2}{\pi M^2}J_2(M(|z|+1/k))\right]\;,
\ee
where $z=\pm (e^{ky}-1)/k\gg 1/k$, $k=1/\ell$, 
and $M/k\ll 1$, and the IR brane is situated at a large $z=z_c$ (and the UV brane is at $z=0$). 
There is a zero mode wave function, the massless graviton, obtained from the $Y_2$ solution in a certain $M\rightarrow 0$ limit, 
and KK modes, all corresponding to states in the boundary field theory. 
The quantization condition for KK modes (QFT states) is obtained by imposing a boundary condition at $z_c$, that amounts to 
\be
Y_1(t+a)+\frac{4}{\pi^2a^2}J_1(t+a)=0\;,\;\; t=M|z_c|\;,\;\; a=M/k=M\ell\ll 1\;,\;\;\; t+a=M_n\ell e^{y_c/\ell}\;,
\ee
so it is approximately again $J_1(t)=J_1(M_n z_c)=0$, but otherwise again involves both $J_1$ and $Y_1$. 
We note that both of these cases give $0^{++}$ QFT states coming from the graviton spectrum in the dual, and 
in the latter case there is also a massless mode. 

In conclusion, we can say that the most general quantization condition is of the type
\be
J_1(M_n \ell K)+c Y_1(M_n\ell K)=0\;,\label{generalquant}
\ee
for some constants $K$ and $c$, and $K$ can be absorbed in the overall rescaling of $M_n$.

\section{Conclusions}

In this paper we have considered the logarithmic (log-divergent) branch of the d'Alembertian operator 
in AdS soliton and related confining holographic backgrounds, we have shown that it is normalizable, and it has the 
interpretation of giving multitrace deformations of the theory, but otherwise not changing the RG flow at the UV, that is a universal 
flow between the UV theory and a fixed IR. 

The one thing that does change is the glueball spectrum, for which we can have different results, depending on the 
deformation. We have considered the deformation with $\a=\lambda^{-1}\b$, which we have found that corresponds to 
a renormalization of the same functional form as the one-loop nonabelian case (\ref{loop}), but now in the IR instead of the UV, 
and we have found that we can use the parameter $\lambda$ to lift the degeneracy of the $0^{++}$ and $2^{++}$ spectra
and match desired lattice data better than previous ones in the literature for $QCD_3$. 

We have also found that we can express to a good degree of accuracy the quantization conditions of the spectra through
a combination of the functions $J_1$ and $Y_1$, as in (\ref{generalquant}), which matches both the results of spectra
from holographic high-energy hadron scattering in QCD, and of spectra from the Randall-Sundrum model, understood as 
AdS/CFT with UV and IR cut-offs. 

The coupling constant that appears due to the IR deformation, $\lambda$, is such that the eigenvalues seem to make a continuos curve on a cylinder. This follows from the fact that $M_{n}^2(\lambda^{-1}=-\infty)=M_{n+1}^2(\lambda^{-1}=\infty)$, and therefore one can identify these points.

The spectrum is free of tachyons provided that the coupling of the deformation satisfies $\lambda^{-1}<\lambda^{-1}_c$. At the onset 
of this instability there is a massless mode and the probe limit is no longer valid. This seems to indicate that the scale at 
which the confining gauge theories confine is important, and that the value of the coupling constant there can be 
constrained by purely theoretical reasoning.

\section*{Acknowledgements}

We thank Carlos N\'{u}\~nez for many detailed comments on the manuscript, and Antal Jevicki, Simon Ross, Guillermo Silva and Jacob 
Sonnenchein for useful comments.
The work of HN is supported in part by  CNPq grant 301491/2019-4 and FAPESP grant 2019/21281-4.
HN would also like to thank the ICTP-SAIFR for their support through FAPESP grant 2021/14335-0.
The work of AA is supported in part by the FAPESP visiting researcher award 2022/11765-7 and the FONDECYT grants 1200986, 1210635, 1221504 and 1230853.

\newpage

\hypersetup{linkcolor=blue}
\phantomsection 
\addtocontents{toc}{\protect\addvspace{4.5pt}}
\bibliographystyle{utphys}
\bibliography{GlueballsFinal}

\end{document}